\documentclass[reprint,notitlepage,onecolumn,superscriptaddress,amsmath,amssymb]{revtex4-1}
\usepackage[utf8]{inputenc}
\usepackage{graphicx}
\usepackage{diagbox}
\usepackage{braket}
\usepackage{multirow}
\usepackage{tabularx}
\usepackage{booktabs}
\usepackage{todonotes}
\usepackage{subfig}
\usepackage{caption}
\usepackage{multirow}
\usepackage{makecell}
\usepackage[ruled,vlined]{algorithm2e}
\usepackage[shortlabels]{enumitem}
\begin{document}
	
\title{Quantum Bayes classifiers and their application in image classification}
\author{Ming-Ming Wang}
\email{bluess1982@126.com}
\author{Xiao-Ying Zhang}
\affiliation{The Shaanxi Key Laboratory of Clothing Intelligence, School of Computer Science, Xi'an Polytechnic University, Xi'an 710048, China}

\begin{abstract}
Bayesian networks are powerful tools for probabilistic analysis and have been widely used in machine learning and data science.
Unlike the time-consuming parameter training process of neural networks, Bayes classifiers constructed on Bayesian networks can make decisions based solely on statistical data from samples.
In this paper, we focus on constructing quantum Bayes classifiers (QBCs). We design both a na\"{\i}ve QBC and three semi-na\"{\i}ve QBCs (SN-QBCs). These QBCs are then applied to image classification tasks.
To reduce computational complexity, we employ a local feature sampling method to extract a limited number of feature attributes from an image. These attributes serve as nodes of the Bayesian networks to generate the QBCs.
We simulate these QBCs on the MindQuantum platform and evaluate their performance on the MNIST and Fashion-MNIST datasets. Our results demonstrate that these QBCs achieve good classification accuracies even with a limited number of attributes. The classification accuracies of QBCs on the MNIST dataset surpass those of classical Bayesian networks and quantum neural networks that utilize all available feature attributes.
\end{abstract}

\date{today}

\maketitle

\section{Introduction}
As a powerful tool for studying causal relationships between variables and inferring the impact of variable states on outcomes,
Bayesian networks are widely used in machine learning and data science, including Monte Carlo analysis \cite{MosegaardSambridge-426}, reliability and risk analysis \cite{ModarresGroth-449}, health monitoring \cite{Daniels-420}, healthcare \cite{GaryJarrett-422}, and biomedical systems \cite{WongLin-428}, etc.
The size of a Bayesian network depends on the number of nodes and their dependencies \cite{Kwisthout-427}.
Learning and inference in complex networks can be challenging, especially when dealing with large-scale Bayesian networks, which have been proven to be an NP-hard problem \cite{ChickeringHeckerman-421}.
The emergence of quantum computing offers a new solution to this challenge.

In recent years, many quantum algorithms have demonstrated quantum supremacy for achieving certain accelerations over their classical counterparts.
For example, the Shor's algorithm \cite{Shor-97} achieves exponential acceleration in solving the large number factorization problem. The Grover's algorithm \cite{Grover-225} achieves quadratic acceleration in searching the unstructured data. In addition, quantum algorithms based on classical machine learning, such as quantum support vector machine \cite{RebentrostMohseni-82} and quantum $K$-nearest neighbor \cite{DangJiang-419}, have also demonstrated quantum accelerations.
Image classification is a fundamental problem in computer vision. With the advancement of quantum machine learning, several quantum classifiers have been developed for image classification, including quantum convolutional neural networks \cite{CongChoi-181,KerenidisLandman-432,HurKim-435,GongPei-455}, quantum $K$-nearest neighbor algorithm \cite{DangJiang-419}, and quantum ensemble methods
\cite{SchuldPetruccione-326,MacalusoClissa-437,AraujoDaSilva-438,MacalusoLodi-431,ZhangWang-409}, etc.
Recent progresses of quantum classifiers could be found in \cite{LiDeng-479}.

Quantum Bayesian networks (QBN) were introduced in 1995 as a simulation of classical ones \cite{Youssef-442}. In 2013, Ozols et al. proposed a quantum version of the rejection sampling algorithm called quantum rejection sampling for Bayesian inference \cite{OzolsRoetteler-445}. In 2016, Moreira and Wichert proposed a quantum-like Bayesian network that uses amplitudes to represent marginal and conditional probabilities \cite{MoreiraWichert-448}. In 2019, Woerner and Egger developed a quantum algorithm \cite{WoernerJ-446} for risk analysis using the principles of amplitude amplification and estimation. Their algorithm can provide a quadratic speed-up compared to classical Monte Carlo methods.
In 2023, Gabriel et al. proposed a quantum algorithm based on quantum walks for quantum Bayesian estimation of gravitational waves parameters from black holes \cite{EscrigCampos-480}.
Recently, Borujeni et al. proposed a quantum circuit representation of Bayesian networks \cite{BorujeniNannapaneni-433}.
They designed quantum Bayesian networks for specific problems, such as stock prediction and liquidity risk assessment.
Walid et al. further proposed an optimized version of quantum circuit for improving the quantum representation of Bayesian networks \cite{FathallahAmor-502}.

Currently, there is a lack of research on quantum Bayes classifiers (QBCs) building on Bayesian networks for solving image classification problems. Different from the parameters learning mode of neural networks, a Bayes classifier makes a classification decisions based only on sample features, without the tedious training process, resulting in lower computational complexity, faster speed, and less resource consumption.
In this paper, we study the construction of quantum Bayes classifiers (QBCs).
Specifically, we design a na\"{\i}ve \cite{RishOthers-411,Leung-429} QBC and three semi-na\"{\i}ve QBCs (SN-QBCs), i.e., the SN-QBC based on SPODE network \cite{Kononenko-450,Zhou-423} with the attribute in the center of an image as the superfather, the SN-QBC based on TAN network \cite{Kononenko-450,Zhou-423}, and the SN-QBC based on symmetric relationship of attributes in an image.
These QBCs are then applied to image classification tasks.
We simulate these QBCs on the MindQuantum platform \cite{Mindquantum-418} and evaluate their performance on the MNIST \cite{LecunBottou-417} and Fashion-MNIST \cite{XiaoRasul-416} datasets.

This paper is organized as follows. Sect. 2 introduces the basic concepts of Bayes classifiers.
Sect. 3 discusses the constructions of QBCs.
Sect. 4 presents the image classification algorithm based on QBCs.
Sect. 5 demonstrates the simulation results of QBCs for image classification on the MNIST and Fashion-MNIST datasets.
Sect. 6 further discusses and concludes the paper.

\section{Bayes classifier}
A Bayes classifier is a statistical classifier based on Bayes' theorem.
It considers selecting the optimal category label based on probabilities and misclassification losses, assuming that all relevant probabilities are known.
Suppose the feature of a sample data is $X=\{x_1, x_2, \cdots, x_i, \cdots, x_n\}$ with $x_i$ is an \textit{attribute} of $X$, and the set of class labels is $Y=\{y_1, y_2, \cdots, y_i, \cdots, y_N\}$. Based on the \textit{posterior probability} $P (y_i|X)$, the expected loss of classifying the sample with feature $X$ as $y_i$ is defined as \cite{Zhou-423}
\begin{equation}\label{eq-2}
R(y_i|X)=\sum_{j=1}^N \lambda_{ij} P(y_j|X),
\end{equation}
where $\lambda_{ij}$ is the misclassification loss. The Bayes classifier attempts to correctly classify new samples with minimal misclassification loss based on the distribution pattern of existing samples. The optimal Bayes classifier can be denoted as
\begin{equation}\label{eq-3}
D(X)=\arg\min\limits_{y\in Y}R(y|X).
\end{equation}

For a specific problem to minimize the misclassification rate, $\lambda_{ij}$ can be written as
\begin{equation}\label{eq-4}
\lambda_{ij}=
\begin{cases}
0, &\text{if}\ i=j;\\
1, &\text{otherwise.}
\end{cases}
\end{equation}
Therefore, the optimal Bayes classifier can be rewritten as
\begin{equation}\label{eq-5}
D(X)=\arg\max\limits_{y\in Y}P(y|X).
\end{equation}
That is, the optimal Bayes classifier selects the class that maximizes the posterior probability $P(y|X)$ given the sample with feature $X$.

Obtaining an accurate posterior probability $P(y|X)$ is critical for a Bayes classifier, but this is often challenging in reality. In the probability framework, the posterior probability $P(y|X)$ can be estimated based on a finite number of training sample. According to Bayes' theorem \cite{Zhou-423}, the posterior probability $P(y|X)$ can be written as
\begin{equation}\label{eq-6}
P(y|X)=\frac{P(X|y)P(y)}{P(X)}
\end{equation}
where $P(X)$ is the evidence factor used for normalization, $P(y)$ is the \textit{class-prior probability}, and $P(X|y)$ is the \textit{class-conditional probability} of the sample $X$ with respect to $y$. According to the law of large numbers \cite{Offmann-RgensenPisier-451}, when there are a sufficient number of independently and identically distributed samples, $P(y)$ can be estimated by the frequency of each class that appears in the training set. As for the class-conditional probability $P(X|y)$, it involves combinations of all attributes in $X$. Assuming each attribute $x_i$ has $d$ possible values, $P(X|y)$ will have $N*d^n$ possible values.

\subsection{Na\"{\i}ve Bayes Classifier}
Clearly, it is difficult to obtain the \textit{class-conditional probabilities} directly from a limited number of training sample, as it will result in the problem of combinatorial explosion in calculation, which becomes more severe with the increase of attributes. The na\"{\i}ve Bayes classifier \cite{RishOthers-411,Leung-429} is based on the assumption of ``independence", which assumes that each attribute independently affects the classification result, as shown in Fig. (\ref{fig-1a}). In this case, Eq. (\ref{eq-6}) can be rewritten as
\begin{equation}\label{eq-7}
P(y|X)=\frac{P(y)\prod_i^n P(x_i|y)}{P(X)},
\end{equation}
where $x_i$ the $i$-th attribute of $X$. Since all $P(X)$ are the same, the na\"{\i}ve Bayes classifier can be represented as
\begin{equation}\label{eq-8}
D_{\text{nb}}=\arg\max\limits_{y\in Y}P(y)\prod_i^n P(x_i|y).
\end{equation}

\subsection{Semi-na\"{\i}ve Bayes Classifier}
The premise of the na\"{\i}ve Bayes classifier is that all attributes satisfy the assumption of independence, but this is often not the case in practical applications. Therefore, the learning method of the semi-na\"{\i}ve Bayes classifier \cite{Kononenko-450,Zhou-423} has emerged, which considers stronger dependencies between attributes while avoids the problem of combinatorial explosion caused by considering the joint probability distribution of all attributes.
\textit{One-Dependent Estimator} (ODE) is the most common strategy for the semi-na\"{\i}ve Bayes classifier, which assumes that each attribute only depends on at most one other attribute besides the label $y$, i.e.,
\begin{equation}\label{eq-9}
P(y|X) \propto P(y)\prod_i^n P(x_i|y, pa_i).
\end{equation}
where $pa_i$ is the dependent attribute (or parent attribute) of $x_i$. The key problem of the semi-na\"{\i}ve Bayes classifier is how to determine the parent attribute of $x_i$.

\begin{figure}
	\centering
	\subfloat[Na\"{\i}ve.]{\label{fig-1a}
		\includegraphics[scale=1]{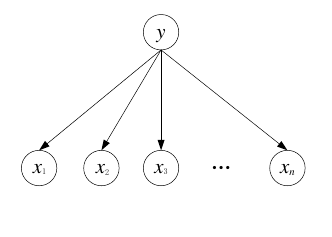}}	
	\subfloat[SPODE.]{\label{fig-1b}
		\includegraphics[scale=1]{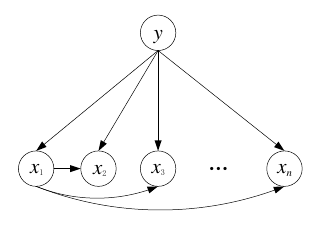}}
	\subfloat[TAN.]{\label{fig-1c}
		\includegraphics[scale=1]{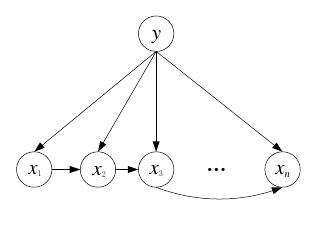}}
\caption{Bayesian networks with different attributes dependencies.}
\label{fig-1}
\end{figure}

A typical approach is to assume that all attributes depend on a single attribute, referred to as the Super-Parent ODE (SPODE), whose dependency relationship is illustrated in Fig. (\ref{fig-1b}).
An alternative method is the Tree Augmented Na\"{\i}ve Bayes (TAN), which calculates the conditional mutual information between any pair of attributes and constructs a maximum weighted tree \cite{Camerini-452,Zhou-423} based on the attribute dependencies, as depicted in Fig. (\ref{fig-1c}).

\section{Quantum Bayes classifiers}
\label{sect-3}
Unlike classical bits that can only represent either 0 or 1, a quantum bit (qubit) can represent both 0 and 1 simultaneously, i.e., quantum superposition.
A single-qubit state can be represented as $\ket{\varphi} = \alpha\ket{0}+\beta\ket{1}$, where $\ket{0}$ and $\ket{1}$ are the basis states of the single-qubit, $\alpha$ and $\beta$ are the amplitudes that satisfy the normalization condition $|\alpha|^2+|\beta|^2 = 1$.
When $\ket{\varphi}$ is measured in the computational basis $\{\ket{0}, \ket{1}\}$, the state will collapse into basis states $\ket{0}$ or $\ket{1}$ with the probability $|\alpha|^2$ or $|\beta|^2$, respectively.
In the quantum gate computing model \cite{NielsenChuang-404}, quantum gates are used to represent unitary operations acting on qubits.
Quantum gates can be divided into single-qubit gates and multi-qubit gates, and any multi-qubit gate can be decomposed into a set of universal quantum gates \cite{NielsenChuang-404}.

In this paper, a quantum circuit for a QBC is constructed using single-qubit gates $R_y$ and $X$, and multi-qubit controlled gates $C^nR_y$. The $X$-gate is a flip gate that flips $\ket{0}$ to $\ket{1}$ or $\ket{1}$ to $\ket{0}$. Its matrix form is
\begin{equation}
X=
\begin{pmatrix}
  1 & 0 \\
  0 & 1 \\
\end{pmatrix}.
\end{equation}
$R_y$ is a single-qubit rotation gate which has the form
\begin{equation}
R_y(\theta)=
\begin{pmatrix}
  \cos\frac{\theta}{2} & -\sin\frac{\theta}{2} \\
  \sin\frac{\theta}{2} & \cos\frac{\theta}{2} \\
\end{pmatrix},
\end{equation}
where $\theta$ is the rotation angle. When acting on $\ket{0}$, the $R_y$-gate generates the following superposition state
\begin{equation*}
R_y (\theta) \ket{0} = \cos\tfrac{\theta}{2}+\sin\tfrac{\theta}{2} \ket{0}.
\end{equation*}
$C^nR_y$ is a multi-qubit controlled rotation gate, where $n$ represents the number of control qubits. When the control qubits are all in $\ket{1}$, the $R_y$ rotation operation is performed on the target qubit. The two-qubit controlled rotation gate $CR_y$ with $n=1$ is represented as
\begin{equation}
CR_y(\theta)=
\begin{pmatrix}
  1 & 0 & 0 & 0 \\
  0 & 1 & 0 & 0 \\
  0 & 0 & \cos\frac{\theta}{2} & -\sin\frac{\theta}{2} \\
  0 & 0 & \sin\frac{\theta}{2} & \cos\frac{\theta}{2} \\
\end{pmatrix}.
\end{equation}
The target qubit $\ket{\varphi}$ will undergo a $R_y(\theta)$ rotation if control qubits are in $\ket{1}$, that is,
\begin{equation*}
C^n R_y (\theta) \ket{1}^{\otimes n}_c \ket{\varphi}_t = \ket{1}^{\otimes n}_c R_y(\theta) \ket{\varphi}_t,
\end{equation*}
where subscript $c$ represents the control qubit and $t$ stands for the target qubit.

\subsection{Na\"{\i}ve-QBC}
\label{sect3-1}
A QBC uses Bayes' rule to perform classification tasks within the framework of quantum computing.
A single-qubit can be used to represent a node in a Bayesian network, and then a superposition quantum state can be used to represent the probability of different label values under various combinations of attributes in the Bayesian network, i.e., $P(y|X)$.

In a na\"{\i}ve Bayesian network, all attributes are independent of each other they depends only on the label. Let $n$ attribute nodes be $x_1$, $x_2$, $\cdots$, $x_n$, and each attribute depends only on the label $y$. The quantum circuit of the na\"{\i}ve-QBC is composed of $n+1$ qubits, corresponding to the label $y$ and $n$ attribute nodes $x_1$ to $x_n$.
The na\"{\i}ve-QBC is constructed as follows:
\begin{enumerate}[(a)]
  \item All $n+1$ qubits are initialized to $\ket{0}$.
  \item
        For the label node $y$, the \textit{class-prior probability} $P(y=0)$ can be obtained by statistically counting the training set. After that, one can encode the class-prior probabilities by $R_y(\cdot)$ operation.
        Unlike the encoding function $\arctan$ used in the Ref. \cite{BorujeniNannapaneni-433}, we use a more natural way of  encoding. That is, let
        \begin{equation}\label{eq-11}
        \cos^2\tfrac{\theta}{2}:=P(y=0).
        \end{equation}
        One can obtain
        \begin{equation}\label{eq-12}
        \theta = 2\arccos\sqrt{P(y=0)}.
        \end{equation}
        This achieves quantum encoding of the \textit{class-prior probabilities}.
        To simplify the representation in the following content, let
        \begin{equation}\label{eq-12-1}
        f(P) := 2\arccos\sqrt{P}.
        \end{equation}

  \item For each attribute node $x_i$, one can get the \textit{class-conditional probabilities} by statistically calculating  $P(x_i=0|y=1)$ and $P(x_i=0|y=0)$. Then, these probabilities are encoded by $CR_y(\cdot)$, where the controlled rotation angles are set as $\theta_{i+1} = f(P(x_i=0|y=1))$ and $\theta_{i+n+1} = f(P(x_i=0|y=0))$, respectively.
\end{enumerate}

\begin{figure}
\centering
\includegraphics[scale=0.7]{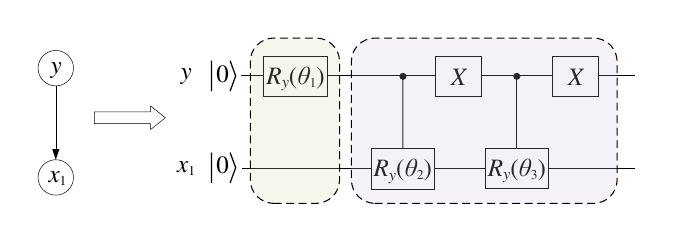}
\caption{The quantum circuit for $y\rightarrow x_1$, i.e., the na\"{\i}ve-QBC with only one attribute ($n=1$).}
\label{fig-2}
\end{figure}

Take the na\"{\i}ve Bayesian network in Fig. \ref{fig-1a} as an example. Suppose these are one label node $y$ and one attribute node $x_1$ ($n=1$), and each node has only 2 values, either 0 or 1.
By statistically counting the \textit{class-prior probabilities} $P(y=0)$ and the \textit{class-conditional probabilities} $P(x_1=0|y=1)$ and $P(x_1=0|y=0)$, one can construct a quantum circuit as depicted in Fig. \ref{fig-2}.
In this case, the output of the circuit is
\begin{equation}\label{eq-13}
\sum_{i,j=\{0,1\}} \sqrt{P(y=i)}\sqrt{P(x_1=j|y=i)}\ket{ij}.
\end{equation}
Note that the probabilities of different label values and attribute values are encoded in the amplitude of the output state, which is the same as Eq. (\ref{eq-8}). That is, the quantum circuit implements the na\"{\i}ve-QBC for features with only one attribute.

\begin{figure}
\centering
\includegraphics[scale=0.7]{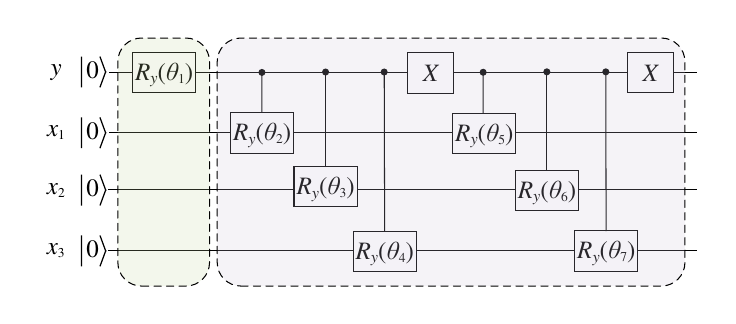}
\caption{The na\"{\i}ve-QBC with 3 attributes. Quantum circuits of QBCs.}
\label{fig-3}
\end{figure}

By continuously adding new attribute nodes to the quantum circuit and establishing controlled rotations between parent nodes and child nodes, one can construct QBCs based on different dependency relationships.
Another example is a Bayesian network with multiple attributes. That is, the na\"{\i}ve QBC with $n=3$ attributes is shown in Fig. \ref{fig-3}. 

In the prediction stage, for a given feature value $X^{\ast}$, one only needs to obtain the probabilities of the basis state $\ket{y=i,X=X^{\ast}}$ by measuring the output state of the na\"{\i}ve QBC. For a binary classification problem, one obtains the values of $P(y=0,X=X^{\ast})$ and $P(y=1,X=X^{\ast})$, and chooses the $y$ value with a higher probability as the classification outcome of the QBC.

\subsection{SPODE-QBC}

For the SPODE Bayesian network, $n$ attribute nodes are $x_1$, $x_2$, $\cdots$, $x_n$. Without loss of generality, assume that $x_1$ is the super-parent node, as shown in Fig. \ref{fig-1b}. The quantum circuit of the SN-QBC based on the SPODE structure (SPODE-QBC) consists of $n+1$ qubits, corresponding to $y$ and $x_1$ to $x_n$. The construction of SPODE-QBC is as follows:
\begin{enumerate}[(a)]
  \item All  $n+1$ qubits are initialized to $\ket{0}$.
  \item For the label node $y$, the \textit{class-prior probability} $P(y=0)$ is counted and encoded by $R_y(\cdot)$, where the rotation angle is set as $\theta_1 = f(P(y=0))$.
  \item For the super-parent $x_1$, one needs to count the \textit{class-conditional probabilities} $P(x_1=0|y=0)$ and $P(x_1=0|y=1)$ and encode these probabilities by using $X$ and $CR_y(\cdot)$, with the controlled rotation angles being set as $\theta_{2} = f(P(x_1=0|y=1))$ and $\theta_{3} = f(P(x_1=0|y=0))$, respectively.
  \item  For the remaining attributes $x_2$ to $x_n$, since each node has two parent nodes, four $C^2R_y(\cdot)$ are used to encode the corresponding \textit{class-conditional probabilities} $P(x_j=0|y, x_1)$, where the controlled rotation angles are set as $ f(P(x_j=0|y, x_1))$ given the values of $y$ and $x_1$.
      That is, when the control bits are $yx_1=00, 01, 10, 11$, the corresponding controlled rotation angles are set as $f(P(x_j=0|y=0, x_1=0))$, $ f(P(x_j=0|y=0, x_1=1))$, $ f(P(x_j=0|y=1, x_1=0))$, and $ f(P(x_j=0|y=1, x_1=1))$, respectively.
\end{enumerate}

\begin{figure}
\centering
\includegraphics[scale=0.7]{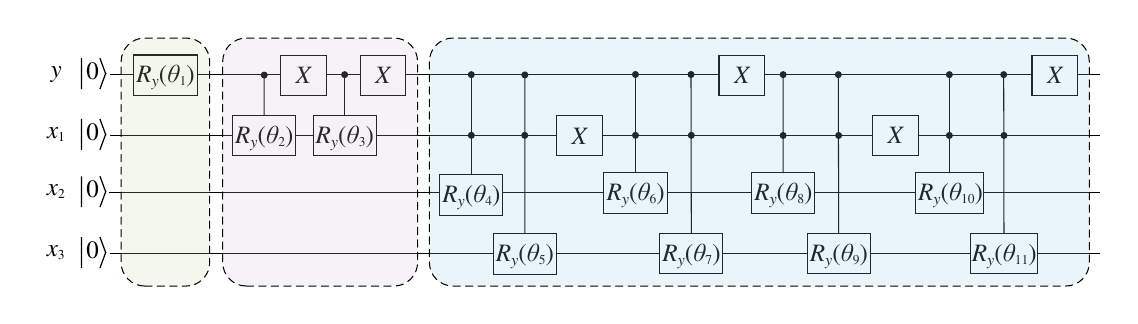}
\caption{The quantum circuit of the SPODE-QBC with 3 attributes and $x_1$ as the super-parent node.}
\label{fig-4}
\end{figure}

For example, for a SPODE structured Bayesian network with 3 attributes shown in Fig. \ref{fig-1b}, the quantum circuit of the SPODE-QBC is depicted in Fig. \ref{fig-4}.
Specifically, the class-conditional probabilities of $x_2$, $x_3$, and $x_4$ are encoded as
$\theta_4=f(P(x_2=0|yx_1=11))$, $\theta_5=f(P(x_3=0|yx_1=11))$,
$\theta_6=f(P(x_2=0|yx_1=10))$, $\theta_7=f(P(x_3=0|yx_1=10))$,
$\theta_8=f(P(x_2=0|yx_1=00))$, $\theta_9=f(P(x_3=0|yx_1=00))$,
$\theta_{10}=f(P(x_2=0|yx_1=01))$, $\theta_{11}=f(P(x_3=0|yx_1=01))$.
The output of the SPODE circuit is
\begin{equation}\label{eq-14}
\sum_{i,j,k,l=\{0,1\}} \sqrt{P(y=i) P(x_1=j|y=i) P(x_2=k|y=i,x_1=j) P(x_3=l|y=i,x_1=j) }\ket{ijkl}.
\end{equation}

\subsection{TAN-QBC}
\label{sect3-3}
For a SN-QBC based on the TAN structure, one needs to obtain the TAN structure Bayesian network firstly. The TAN structure Bayes classifier is generated based on the maximum-weighted spanning tree algorithm \cite{Camerini-452,Zhou-423}, which includes the following steps \cite{Zhou-423}:
\begin{enumerate}[(1)]
  \item  Calculate the conditional mutual information between any two nodes $x_i$ and $x_j$ using the following equation
        \begin{equation}\label{eq-15}
        I(x_i, x_j|y):=\sum_{x_i, x_j, c \in Y} P(x_i, x_j|c) \log\frac{P(x_i,x_j |c)}{P(x_i |c)P(x_j |c)}.
        \end{equation}
  \item  Build a complete graph among nodes and set $I(x_i, x_j|y)$  as the weight between $x_i$ and $x_j$.
  \item  Construct the maximum-weighted spanning tree of the complete graph and set the direction of each edge outward from the root.
  \item  Add the label node $y$ and directed edges from $y$ to each node.
\end{enumerate}

For a Bayesian network with a TAN structure, the quantum circuit of the TAN-QBC consists of $n+1$ qubits, corresponding to $y$ and $x_1$ to $x_n$. The construction of TAN-QBC is as follows:
\begin{enumerate}[(a)]
  \item All qubits are initialized to $\ket{0}$.
  \item The \textit{class-prior probability} $P(y=0)$ is encoded by $R_y(f(P(y=0)))$.
  \item Starting from the root node of the feature spanning tree, the \textit{class-conditional probabilities} of each node are encoded layer by layer.
      For attribute $x_j$, it has at most two parent nodes, i.e., $y$ and $x_{\text{parent}_j}$, where $x_{\text{parent}_j}$ is the parent node of $x_j$ on the upper layer (note that the root node of the feature spanning tree only has one parent $y$).
      The \textit{class-conditional probabilities} $P(x_j|y, x_{\text{parent}_j})$ are encoded into the circuit as the rotation angles of four $C^{2}R_y(\cdot)$ gates.
\end{enumerate}

\begin{figure}
	\centering
	\subfloat[The TAN structure Bayesian network.]{\label{fig-5a}
		\includegraphics[scale=0.8]{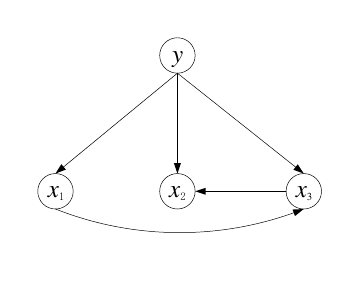}}\\	
	\subfloat[The circuit of the TAN-QBC.]{\label{fig-5b}
		\includegraphics[scale=0.7]{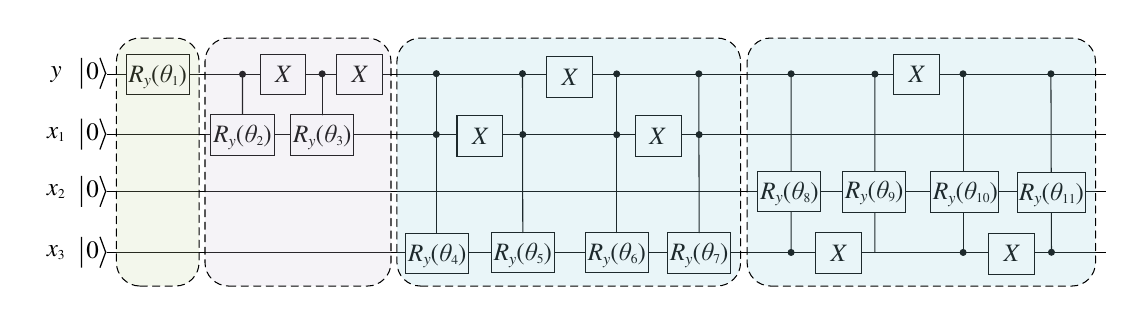}}
\caption{The quantum circuit of the TAN-QBC based on the TAN structure Bayesian network.}
\label{fig-5}
\end{figure}

A simple example of a Bayesian network with a TAN structure is given in Fig. \ref{fig-5a}.
While the circuit of the TAN-QBC based on the TAN structure is presented in Fig. \ref{fig-5b}.
The rotation angles of the label node $y$ and the root node $x_1$ of the spanning tree are set as the na\"{\i}ive-QBCs shown in Sect. \ref{sect3-1}.
The rotation angles of $C^2R_y$ gates acting on $x_3$ are set as
$\theta_4=f(P(x_3=0|yx_1=11))$, $\theta_5=f(P(x_3=0|yx_1=10))$, $\theta_6=f(P(x_3=0|yx_1=00))$,  and $\theta_7=f(P(x_3=0|yx_1=01))$, respectively.
While the rotation angles of $C^2R_y$ gates acting on $x_2$ are set as
$\theta_8=f(P(x_2=0|yx_3=01))$, $\theta_9=f(P(x_2=0|yx_3=00))$, $\theta_{10}=f(P(x_2=0|yx_3=10))$, and $\theta_{11}=f(P(x_2=0|yx_3=11))$, respectively.
The output of the TAN structure SN-QBC is
\begin{equation}\label{eq-16}
\sum_{i,j,k,l=\{0,1\}} \sqrt{P(y=i) P(x_1=j|y=i) P(x_3=k|y=i,x_1=j)P(x_2=l|y=i,x_3=k) }\ket{ijkl}.
\end{equation}

\subsection{SN-QBC Based on the Symmetric Relationship of Image Attributes}
\begin{figure}
\centering
\includegraphics[scale=0.8]{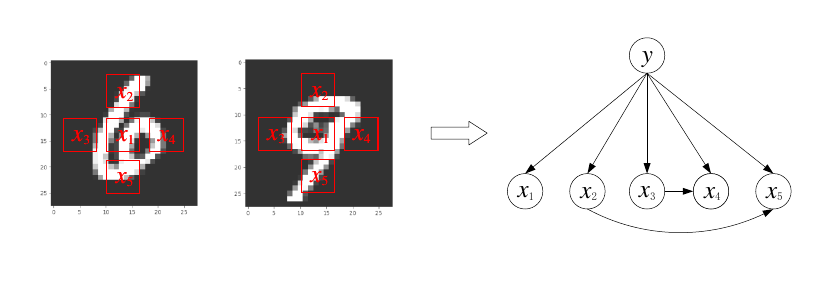}	
\caption{The Bayesian network based on symmetric relationships of feature's attributes.}
\label{fig-8}
\end{figure}

Both the na\"{\i}ve Bayes classifier and the semi-na\"{\i}ve Bayes classifier only consider the dependencies between a small number of attribute nodes.
Note that there are symmetric relationships among the sampled feature attributes in some images, such as the digits ``6" and ``9" shown in Fig. \ref{fig-8}, where there exists symmetric relationships between attributes $x_2$ and $x_5$, and $x_3$ and $x_4$.
These symmetric relationships of image attributes can be used to build Bayesian networks, which can then be used to establish corresponding SN-QBCs.
The method for constructing a Bayesian network based on the symmetric relationships of image attributes is as follows.
\begin{enumerate}[(1)]
  \item Establish a na\"{\i}ve Bayesian network as shown in Fig. \ref{fig-1a};
  \item Consider the symmetric relationship of attributes in the sample images and add a directed edge between each pair of attributes that are symmetrical to each other.
\end{enumerate}
For example, for an image dataset of digits ``6" and ``9", a Bayesian network can be established as shown in Fig. \ref{fig-8}.

For the Bayesian network with symmetric relationships between attributes, in the case of not considering the label node $y$, the network will consist of several independent trees.
In this case, one can construct the corresponding symmetric-QBC by using the method for the TAN-QBC in Sect. \ref{sect3-3}. That is,
\begin{enumerate}[(a)]
  \item Each node is initialized to $\ket{0}$.
  \item The \textit{class-prior probability} $P(y=0)$ is encoded by $R_y(\cdot)$.
  \item Each independent tree is considered separately. For each tree, starting from the root node, the \textit{class-conditional probabilities} of each node are encoded by $CR_y(\cdot)$ layer by layer, respectively.
\end{enumerate}

\section{Image Classification Based on QBCs}

\begin{figure}	
\centering
\includegraphics[scale=0.8]{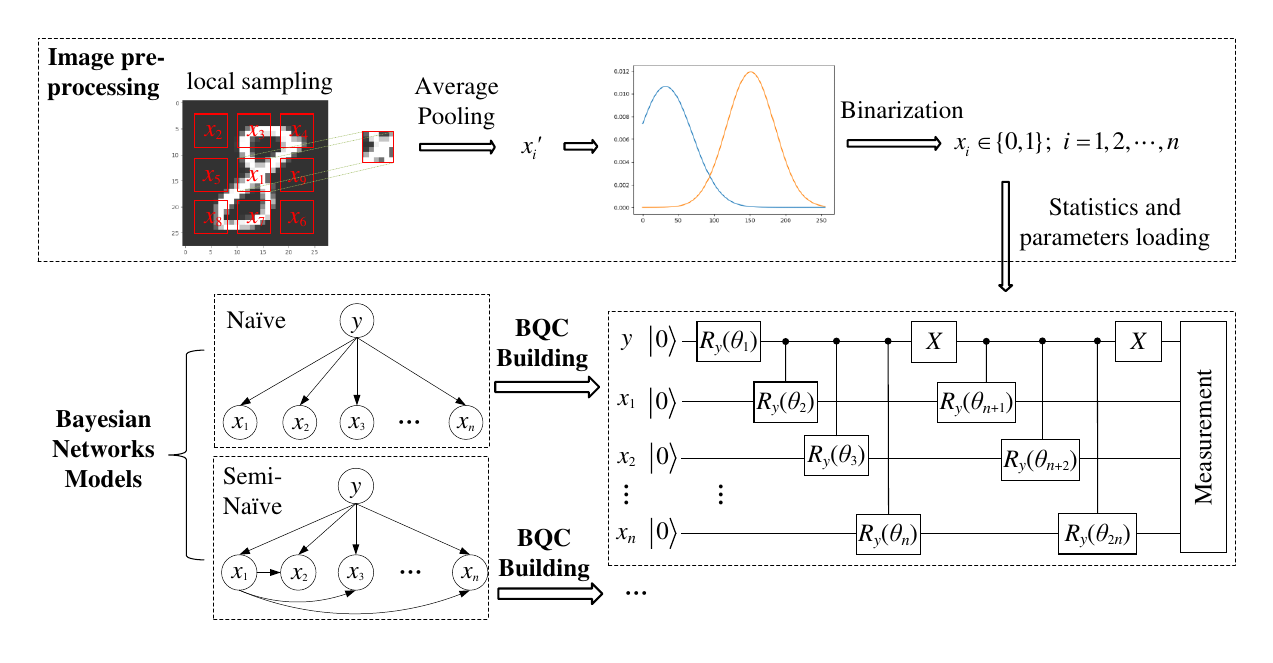}
\caption{The framework of image classifications based on QBCs.}
\label{fig-6}
\end{figure}

The key to the accuracy of a Bayesian classifier lies in feature selection and the structure of Bayesian networks.
In this paper, we propose an image classification framework based on QBCs and local feature sampling, as illustrated in Fig. (\ref{fig-6}).
Firstly, some local areas are selected in an image for feature sampling. Then, the sampled pixels are pooled and binarized to obtain local binary attributes of features. Finally, a QBC can be constructed based on a Bayesian network model with local attributes as nodes for image classification.

\subsection{Local Feature Sampling}
\label{sect4-1}
For image classifications using a na\"{\i}ve QBC, if each pixel of the image is taken as a node of a Bayesian network, the network will be very complex. For example, a 28*28 image requires 784 nodes to represent the network.
Although $n$ qubits can form a feature space with a dimension of $2^n$, the resources required for the quantum Bayesian networks are still relatively large.
In addition, considering the complex relationship of interdependence among feature's attributes, the complexity of Bayesian networks will continue to increase, which hinders the implementation of QBCs on current NISQ devices.

To reduce computational complexity and utilize features effectively for classification, we propose the local feature sampling method. This method aims to obtain a small number of local key attributes from an image.
For image classification, background pixels shared by some images might not provide useful information for accurate classification by Bayes classifiers.
The local feature sampling method can reduce the number of attribute nodes in Bayesian networks, decrease the scale and computational complexity of QBCs, and diminish the influence of quantum noise, which is beneficial for the experimental implementation of quantum circuits.

\begin{figure}
	\centering
	\subfloat[One intersection.]{\label{fig-7a}
		\includegraphics[scale=0.3]{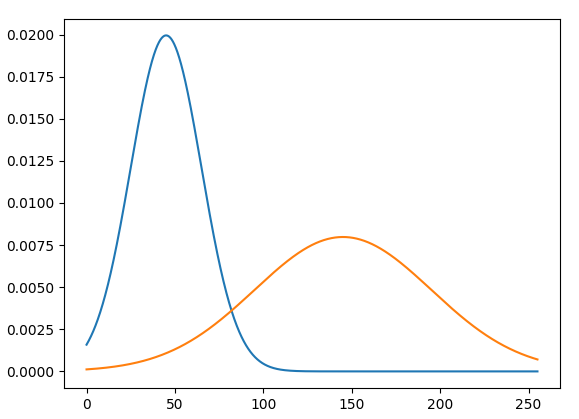}}	
	\subfloat[Two intersections.]{\label{fig-7b}
		\includegraphics[scale=0.3]{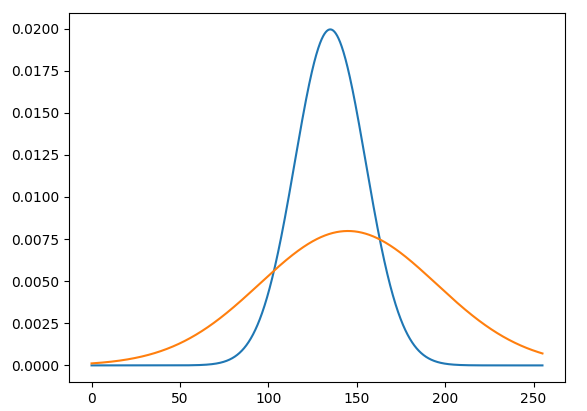}}
\caption{Two Gaussian functions with intersections.}
\label{fig-7}
\end{figure}

As for data preprocessing, the local feature sampling method is illustrated in Fig. \ref{fig-6}, which is performed as follows.
\begin{enumerate}[(1)]
  \item
  \textbf{Local sampling and average pooling}.
  Using a convolution operation on the sampling block and the convolution kernel, an image is locally sampled according to the specified block size (convolution kernel size).
  Next, the appropriate average pooling value $x'_i$ of the attribute node $x_i$ in the Bayesian network is obtained by applying average pooling \cite{YuWang-424}.

  \item
  \textbf{Feature binarization}.
  In a classical Bayes classifier, assume that the probability density function $p(x_i |y)\sim N(\mu_{y,i},\sigma_{y,i}^2)$, where $\mu_{y,i}$ and $\sigma_{y,i}^2$ are the mean and variance of the attribute $x_i$ on the $y$-th class.
  One can obtain the \textit{class-prior probability} by calculating the probability density function.
  However, it is challenging to replicate this process for a QBC.
  In this case, binarization is used to transform $x_i$ into 0 or 1 such that it can be represented by a single-qubit.

  Here, we use the maximum likelihood estimation (MLE) \cite{Dutilleul-453,BoullE-425} method adopted in classical Bayes classifiers for obtaining \textit{class-conditional probabilities} of continuous variables.
  For the binary classification problem, this method runs as follows.
  Assume that $p(x_i |y_0)\sim N(\mu_{y_0,i},\sigma_{y_0,i}^2)$ and $p(x_i |y_1)\sim N(\mu_{y_1,i},\sigma_{y_1,i}^2)$.
  When two Gaussian functions have only one intersection, which is denoted as $x_{\text{ins}}$, as shown in Fig. (\ref{fig-7a}), the attribute value is set as follows
    \begin{equation}\label{eq-17}
    x_i =
    \begin{cases}
    0, &\text{if}\  x'_i \leq x_{\text{ins}}\ \text{and}\ \mu_{y_0,i}\leq\mu_{y_1,i},\ \text{or}\ x'_i >x_{\text{ins}}\ \text{and}\  \mu_{y_0,i}>\mu_{y_1,i};\\
    1, &\text{otherwise.}
    \end{cases}
    \end{equation}
    where $x'_i$ is the average pooling value generated by the the average pooling processing.
    If there are two intersections $x_{\text{ins1}}$ and $x_{\text{ins2}}$ with $x_{\text{ins1}} \leq x_{\text{ins2}}$, as shown in Fig. (\ref{fig-7b}), in cases where $\mu_{y_0,i} \leq \mu_{y_1,i}$, the attribute value is set as
    \begin{equation}\label{eq-18}
    x_i =
    \begin{cases}
    0, &\text{if}\ x'_i \leq x_{\text{ins1}},\ \text{or}\ x_{\text{ins1}} \leq x'_i \leq x_{\text{ins2}}\ \text{and}\  |x_{\text{ins1}}-x'_i| \leq |x_{\text{ins2}}-x'_i|;\\
    1, &\text{otherwise.}
    \end{cases}
    \end{equation}
    Conversely, when $\mu_{y_0,i} > \mu_{y_1,i}$, the attribute value is set as
    \begin{equation}\label{eq-18}
    x_i =
    \begin{cases}
    1, &\text{if}\ x'_i \leq x_{\text{ins1}},\ \text{or}\ x_{\text{ins1}}\leq x'_i \leq x_{\text{ins2}}\ \text{and}\  |x_{\text{ins1}}-x'_i|\leq |x_{\text{ins2}}-x'_i|;\\
    0, &\text{otherwise.}
    \end{cases}
    \end{equation}

\end{enumerate}

By utilizing the aforementioned method, it is possible to obtain a limited number of local attributes of features.
These local attributes serve as nodes of Bayesian networks for the constructions of QBCs.

\subsection{Image Classification Algorithm Based on QBCs}
The image classification algorithm based on QBCs and local feature sampling is shown in Fig. (\ref{fig-6}).
The algorithm consists of two stages, namely the image preprocessing stage, and the Bayesian networks and QBCs construction stage.
In the image preprocessing stage, certain local areas are chosen for sampling.
After that, the sampled attributes are pooled, and the Gaussian method is used for binarization. This process converts the sampled key attributes into either 0 or 1, allowing an attribute to be represented by a single-qubit in QBCs.
In the second stage, a Bayesian network model is selected and the corresponding QBC is constructed with local key attributes as nodes.
The algorithm runs as follows.

\begin{enumerate}[(a)]
  \item A Bayesian network is selected. 
  \item \label{step-2}
  Local sampling and average pooling are performed on images from the training set, and the mean value $\mu_i$ and variance $\sigma_i$ of the corresponding attributes are calculated. The Gaussian binarization method is executed to obtain the binarized attributes, as described in Sect. \ref{sect4-1}.
  \item The QBC circuit is constructed based on the chosen Bayesian network model by using local binarized attributes as nodes. Also see Sect. \ref{sect-3}.
  \item  The \textit{class-prior probability} and \textit{class-conditional probability} required for the Bayesian network are calculated statistically. These values are then loaded into the QBC circuit as controlled angles to complete the construction of the QBC.
  \item To predict the class of a new image, one needs to repeat the step \ref{step-2} to obtain a new local key feature $X'$ of the image, and then measures the probabilities of states $\ket{0X'}$ and $\ket{1X'}$ on the QBC circuit. The class with the highest probability will be the classification result of the image.
\end{enumerate}

\section{Simulations}
\begin{figure}
\centering
\includegraphics[scale=0.8]{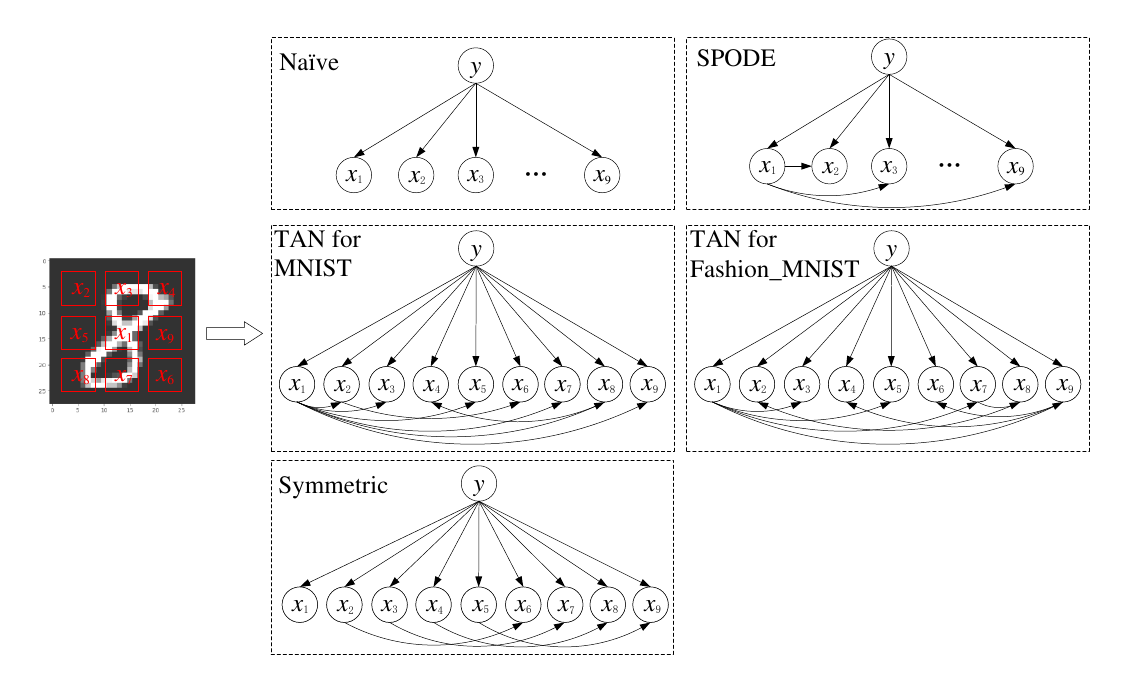}
\caption{The sampling points and four types of Bayesian networks used in the simulation, i.e., the na\"{\i}ve, the SPODE, the TAN, and the symmetric Bayesian networks.}
\label{fig-9}
\end{figure}

In this paper, we simulate four QBCs on the MindQuantum \cite{Mindquantum-418} quantum simulation platform. And the performances of these QBCs are tested using the MNIST \cite{LecunBottou-417} and Fashion-MNIST \cite{XiaoRasul-416} datasets.
In the preprocessing stage, the sampling block size is set to 7*7, and the average pooling method is applied.
All images are sampled according to the sampling positions shown in Fig. (\ref{fig-9}) to obtain 9 local key attributes ($n=9$).

As is shown in Fig. (\ref{fig-9}), four Bayesian network models are used to construct four corresponding QBCs, i.e., the na\"{\i}ve QBC, the SN-QBC based on the SPODE structure (SPODE-QBC), the SN-QBC based on the TAN structure (TAN-QBC), and the SN-QBC based on the symmetric relationships of attributes (symmetric-QBC).
For the TAN structures of the MNIST and Fashion-MNIST datasets, the spanning trees of the training data of the two datasets are calculated according to the maximum weighted spanning tree algorithm \cite{Camerini-452,Zhou-423} introduced in Sect. \ref{sect3-3}.

\subsection{The Accuracy of Binary Classification}
\begin{figure}
\centering
\includegraphics[scale=0.4]{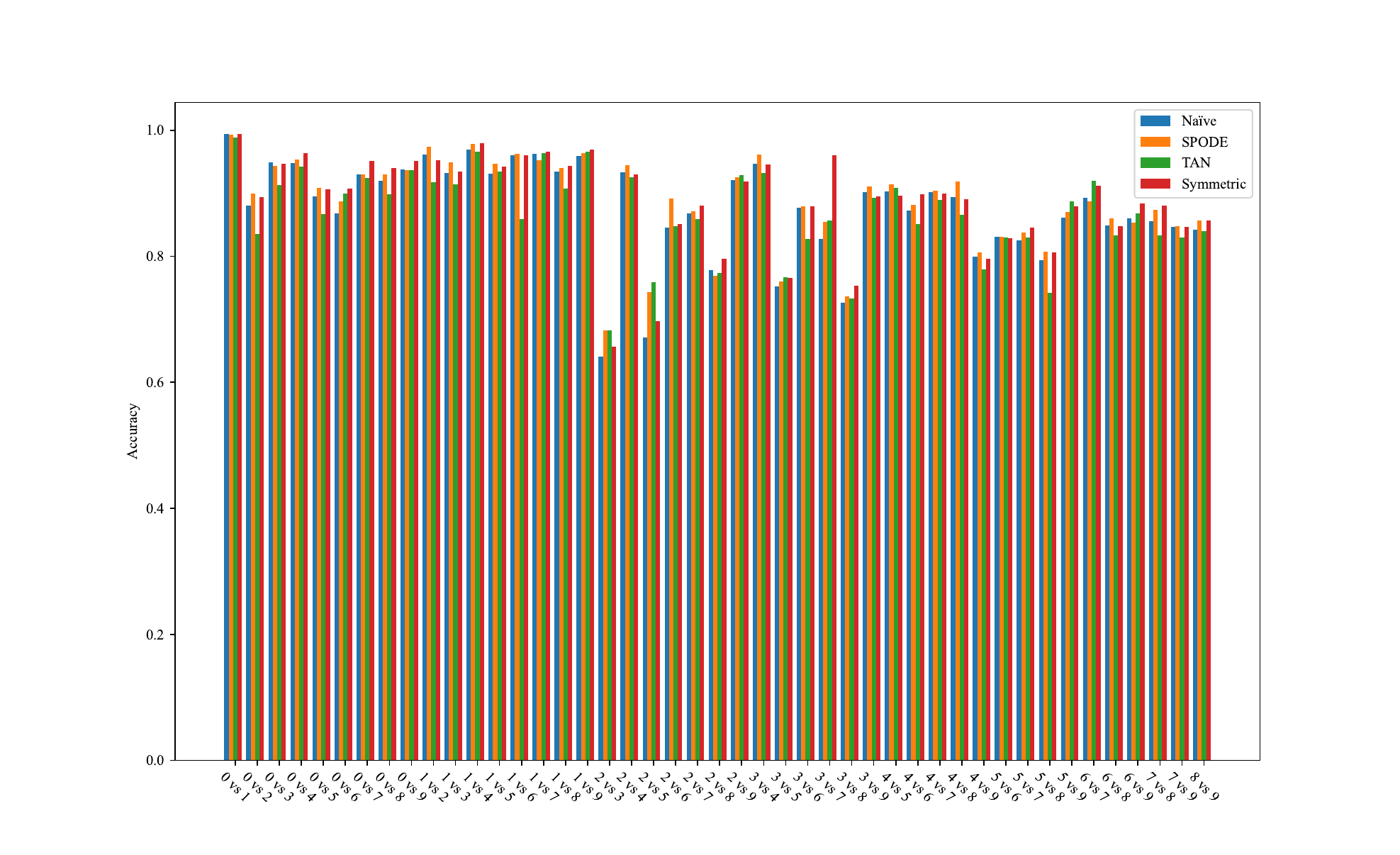}
\caption{The classification accuracies of every two classes in the MNIST dataset on the na\"{\i}ve QBC, the SPODE-QBC, the TAN-QBC, and the symmetric-QBC.}
\label{fig-11}
\end{figure}

\begin{figure}
\centering
\includegraphics[scale=0.4]{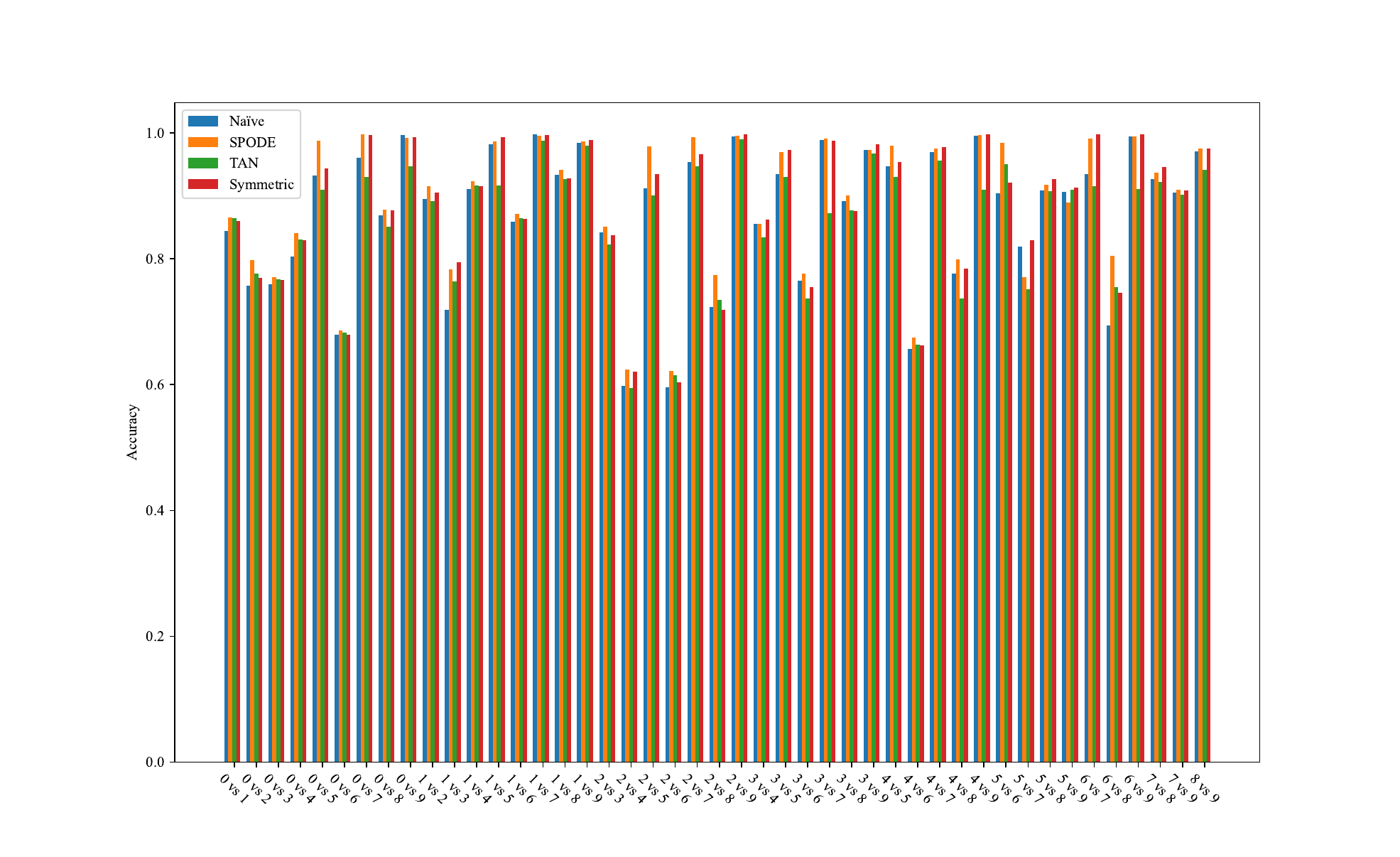}
\caption{The classification accuracies of every two classes in the Fashion-MNIST dataset on the na\"{\i}ve QBC, the SPODE-QBC, the TAN-QBC, and the symmetric-QBC.}
\label{fig-12}
\end{figure}

The MNIST and Fashion-MNIST datasets are used to verify the binary classification effects of four QBCs on every two image classes.  Fig. \ref{fig-11} and Fig. \ref{fig-12} show the classification accuracies of the four QBCs in the MNIST and the Fashion-MNIST datasets, respectively.

It can be seen from Fig. \ref{fig-11} that the symmetric-QBC shows a better classification performance in most cases.
While as is shown in Fig. \ref{fig-12}, the SPODE-QBC shows better results in the Fashion-MNIST dataset in most cases, and the symmetric-QBC also exhibits better classification results on some data.

In addition, the simulation results show that QBCs still does not achieve ideal classification results for some data in these two datasets, which lies in the following reasons.
\begin{enumerate}[(1)]
  \item The Bayesian network used in the simulation may not be suitable for all data. Different Bayesian networks should be considered for different data;
  \item The hyperparameters involved in the simulation, such as the sampling block size, the convolution kernel size, and the pooling method, etc., has a significant impact on the results, which should be selected carefully;
  \item Some attributes extracted from the local sampling and binarization procedure can not distinguish data from two classes very well. Other sampling and binarization methods could be considered.
\end{enumerate}

\subsection{Overall Performance of QBCs}
To evaluate the overall performance of each QBC, the average classification accuracy, variance, average precision, average recall, and average $F_1$ score \cite{Liu-408,Caelen-407} are calculated.
The average classification accuracy of a binary classifier for two classes in the MNIST and Fashion-MNIST datasets is defined as
\begin{equation}\label{eq-20}
    \overline{acc}=\frac{1}{45}\sum_{i=0}^{8}\sum_{j=i+1}^{9}acc_{ij},
\end{equation}
where $acc_{ij}$ represents the accuracy of binary classification of classes $i$ and $j$.
The variance is defined as
\begin{equation}\label{eq-21}
    s^2=\frac{1}{45}\sum_{i=0}^{8}\sum_{j=i+1}^{9}(\overline{acc}-acc_{ij})^2.
\end{equation}
While the average precision, average recall, and average $F_1$ score are defined as
\begin{equation}\label{eq-22}
    \overline{prec}=\frac{1}{45}\sum_{i=0}^{8}\sum_{j=i+1}^{9}\frac{TP_{ij}}{TP_{ij}+FP_{ij}},\quad
    \overline{recall}=\frac{1}{45}\sum_{i=0}^{8}\sum_{j=i+1}^{9}\frac{TP_{ij}}{TP_{ij}+FN_{ij}},\quad
    F_1=\frac{2*\overline{prec}*\overline{recall}}{\overline{prec}+\overline{recall}},
\end{equation}
where TP means the number of true positive, FP means false positive, and FN means false negative.
Table \ref{table-4} shows the overall performance of the four QBCs mentioned above for all binary classification pairs in the MNIST and Fashion-MNIST datasets.
As is shown in Table \ref{table-4}, the SPODE-QBC and symmetric-QBC exhibit good classification accuracies in both the MNIST and Fashion-MNIST datasets, while the na\"{\i}ve-QBC and TAN-QBC also show relatively good classification effects.
On the whole, the symmetric-QBC performs best in the MNIST dataset, while the SPODE-QBC performs best in the Fashion-MNIST dataset.

\begin{table}[]
\caption{Overall performances of the four QBCs in two datasets.}\label{table-4}
\begin{tabular}{llccccc}
\hline\\
        Dataset       & Classifier   & $\overline{acc}$   & $s^2$   & $\overline{precision}$  & $\overline{recall}$  & $F_1$ \\
\\\hline
\multirow{4}{*}{MNIST} &
      Na\"{\i}ve-QBC  & 0.8767 & 0.0777 & 0.8844 & 0.8689 & 0.8762 \\
&   SPODE-QBC     & 0.8873 & \textbf{0.0707} & \textbf{0.8930 }& 0.8808 & 0.8864 \\
&   TAN-QBC          & 0.8055 & 0.0888 & 0.7390 & \textbf{0.9775} & 0.8388 \\			 	
&   Symmetric-QBC & \textbf{0.8889} &0.0739 & \textbf{0.8930} & 0.8813 & \textbf{0.8867}\\
\hline
\multirow{4}{*}{Fashion-MNIST} &
      Na\"{\i}ve-QBC  & 0.8712 & 0.1128 & 0.8726 & 0.8748 & 0.8728 \\
&   SPODE-QBC     & \textbf{0.8916} & \textbf{0.1080} & \textbf{0.8863} & 0.9107  & \textbf{0.8962} \\
&   TAN-QBC          & 0.8739 &	0.1115   &	 0.8343  &	\textbf{0.9602} &	0.8894\\
&   Symmetric-QBC   & 0.8834   & 0.1112 & 0.8807 & 0.8942 & 0.8859 \\
\hline
\end{tabular}
\end{table}

\subsection{Comparison with Other Classifiers}
Table \ref{table-5} shows the comparison of classification performance of the classical
Gaussian na\"{\i}ve Bayes classifier \cite{GaussianNB}, the quantum convolutional neural network (QCNN) \cite{HurKim-435}, and four QBCs for classes 0 and 1 in the MNIST and Fashion-MNIST datasets.
The results show that QBCs performs better than the classical Bayes classifier and QCNN in the MNIST dataset, while the QCNN performs best in the Fashion-MNIST dataset. It is important to note that the classical Bayesian classifier and quantum neural network classifier use all 786 features, while the four QBCs in this paper use only 9 binary features.
The simulation also shows that adjusting the sampling block size and the convolution kernels can improve the accuracy of QBCs, but this may not be effective for all classes of images. Specific analysis and adaptation are required for a specific task.

\begin{table}[]
\caption{Comparison of classification accuracy of classes 0 and 1 in MNIST and Fashion-MNIST by different classifiers.}\label{table-5}
\begin{tabular}{ccc}
\hline\\
        Classifier    & \makecell[c]{MNIST\\0 vs 1}   & \makecell[c]{Fashion-MNIST:\\0 (t-shirt) vs 1 (trouser)} \\
\\\hline
Classical na\"{\i}ve Bayes \cite{GaussianNB}  & 0.985	& 0.897 \\
QCNN\cite{HurKim-435}                & 0.987 &	\textbf{0.941}\\
Na\"{\i}ve-QBC                                 & \textbf{0.994}	& 0.844 \\
SPODE-QBC                                       & 0.992	& 0.866 \\
TAN-QBC                                            & 0.968 & 	0.820 \\
Symmetric-QBC                                 & \textbf{0.994}	& 0.861 \\
\hline
\end{tabular}
\end{table}

\section{Discussions \& Conclusions}
In this paper, we study the constructions of QBCs using Bayesian networks. Based on four kinds of Bayesian networks, four QBCs are implemented, that is, the na\"{\i}ve-QBC, the SPODE-QBC with the attribute in an image center as the `superparent', the TAN-QBC, and the symmetric-QBC.
We apply these QBCs to image classification and propose the image classification algorithm.
To reduce the computational complexity, a local feature sampling method is designed to extract a few key attributes from a huge number of image pixels.
By retaining essential attributes, this method achieves high classification accuracies while reducing the feature size.
Adjusting the sampling block size and utilizing different convolution kernels can further improve the accuracies of QBCs.

The classification effects of QBCs are verified on the MindQuantum platform for the MNIST and Fashion-MNIST datasets, respectively. The results show that QBCs perform well for image classification.
We also compare QBCs with the classical Bayesian classifier and QCNN that use much larger size of features, which shows that QBCs using a fewer size of features still have advantages in some cases.
Unlike QCNN classifiers \cite{HurKim-435,CongChoi-181}, QBCs do not require a time-consuming training process. They only require to calculate the corresponding parameters statistically and load them into quantum circuits, while the classification decision can be made by measuring quantum circuits, which is faster, lower in computational complexity, and less resource-consuming than QCNN.
At present, there does not exist a general QBC model that performs well with all types of data. Building QBCs based on alternative varieties of Bayesian networks that work well with various datasets would be intriguing.

\section*{Data availability statement}
The data that support the findings of this study are available from the corresponding author upon reasonable request.

\section*{Acknowledgements}
This project was supported by the National Natural Science Foundation of China (Grant No. 61601358).


\begin{thebibliography}{10}
\providecommand{\url}[1]{{#1}}
\providecommand{\urlprefix}{URL }
\expandafter\ifx\csname urlstyle\endcsname\relax
  \providecommand{\doi}[1]{DOI \discretionary{}{}{}#1}\else
  \providecommand{\doi}{DOI \discretionary{}{}{}\begingroup
  \urlstyle{rm}\Url}\fi

\bibitem{MosegaardSambridge-426}
K.~Mosegaard, M.~Sambridge, Monte carlo analysis of inverse problems, Inverse
  Probl., 2002, 18(3), R29

\bibitem{ModarresGroth-449}
M.~Modarres, K.~Groth, \emph{Reliability and risk analysis} (CRC Press, 2023)

\bibitem{Daniels-420}
N.~Daniels, Justice, health, and healthcare, Am. J. Bioeth., 2001, 1(2), 2

\bibitem{GaryJarrett-422}
P.~Gary~Jarrett, Logistics in the health care industry, Int. J. Phys. Distrib.
  Logist. Manag., 1998, 28(9/10), 741

\bibitem{WongLin-428}
H.~Wong, W.~Lin, H.~Laure, Multi-polarization reconfigurable antenna for
  wireless biomedical system, IEEE Trans. Biomed. Circuits Syst., 2017, 11(3),
  652

\bibitem{Kwisthout-427}
J.~Kwisthout, Most probable explanations in bayesian networks: Complexity and
  tractability, Int. J. Approx. Reason., 2011, 52(9), 1452

\bibitem{ChickeringHeckerman-421}
M.~Chickering, D.~Heckerman, C.~Meek, Large-sample learning of bayesian
  networks is np-hard, J. Mach. Learn. Res., 2004, 5, 1287

\bibitem{Shor-97}
P.W. Shor, Polynomial-time algorithms for prime factorization and discrete
  logarithms on a quantum computer, SIAM J. Sci. Statist. Comput., 1997, 26(5),
  1484

\bibitem{Grover-225}
L.K. Grover, Quantum mechanics helps in searching for a needle in a haystack,
  Phys. Rev. Lett., 1997, 79(2), 325

\bibitem{RebentrostMohseni-82}
P.~Rebentrost, M.~Mohseni, S.~Lloyd, Quantum support vector machine for big
  data classification, Phys. Rev. Lett., 2014, 113(13), 130503

\bibitem{DangJiang-419}
Y.~Dang, N.~Jiang, H.~Hu, Z.~Ji, W.~Zhang, Image classification based on
  quantum k-nearest-neighbor algorithm, Quantum Inf. Process., 2018, 17, 1

\bibitem{CongChoi-181}
I.~Cong, S.~Choi, M.D. Lukin, Quantum convolutional neural networks, Nat.
  Phys., 2019, 15(12), 1273

\bibitem{KerenidisLandman-432}
I.~Kerenidis, J.~Landman, A.~Prakash, Quantum algorithms for deep convolutional
  neural networks, arXiv preprint arXiv:1911.01117, 2019

\bibitem{HurKim-435}
T.~Hur, L.~Kim, D.K. Park, Quantum convolutional neural network for classical
  data classification, Quantum Mach. Intell., 2022, 4(1), 3

\bibitem{GongPei-455}
L.H. Gong, J.J. Pei, T.F. Zhang, N.R. Zhou, Quantum convolutional neural
  network based on variational quantum circuits, Opt. Commun., 2024, 550,
  129993

\bibitem{SchuldPetruccione-326}
M.~Schuld, F.~Petruccione, Quantum ensembles of quantum classifiers, Sci. Rep.,
  2018, 8(1), 2772

\bibitem{MacalusoClissa-437}
A.~Macaluso, L.~Clissa, L.~Stefano, C.~Sartori, Quantum ensemble for
  classification, arXiv preprint arXiv:2007.01028, 2020

\bibitem{AraujoDaSilva-438}
I.C. Araujo, A.J. Da~Silva, Quantum ensemble of trained classifiers, in
  \emph{2020 International Joint Conference on Neural Networks (IJCNN)} (IEEE,
  Glasgow, UK), 2020, pp. 1--8

\bibitem{MacalusoLodi-431}
A.~Macaluso, S.~Lodi, C.~Sartori, Quantum algorithm for ensemble learning, in
  \emph{21st Italian Conf. Theoretical Computer Science (CEUR-WS.org, 2020)}
  2020, pp. 149--154

\bibitem{ZhangWang-409}
X.Y. Zhang, M.M. Wang, An efficient combination strategy for hybrid quantum
  ensemble classifier, Int. J. Quantum Inf., 2023, 21(06), 2350027

\bibitem{LiDeng-479}
W.~Li, D.L. Deng, Recent advances for quantum classifiers, Sci. China Phys.
  Mech. Astron., 2021, 65(2), 220301

\bibitem{Youssef-442}
S.~Youssef, Quantum mechanics as an exotic probability theory, arXiv preprint
  quant-ph/9509004, 1995

\bibitem{OzolsRoetteler-445}
M.~Ozols, M.~Roetteler, J.R.M. Roland, Quantum rejection sampling, ACM Trans.
  Comput. Theory, 2013, 5(3), 1

\bibitem{MoreiraWichert-448}
C.~Moreira, A.~Wichert, Quantum-like bayesian networks for modeling decision
  making, Front. Psychol., 2016, 7, 11

\bibitem{WoernerJ-446}
S.~Woerner, D.J. Egger, Quantum risk analysis, npj Quantum Inf., 2019, 1(5), 15

\bibitem{EscrigCampos-480}
G.~Escrig, R.~Campos, P.A.M. Casares, M.A. Martin-Delgado, Parameter estimation
  of gravitational waves with a quantum metropolis algorithm, Class. Quantum
  Grav., 2023, 40(4), 045001

\bibitem{BorujeniNannapaneni-433}
S.E. Borujeni, S.~Nannapaneni, N.H. Nguyen, E.C. Behrman, J.E. Steck, Quantum
  circuit representation of bayesian networks, Expert Syst. Appl., 2021, 176,
  114768

\bibitem{FathallahAmor-502}
W.~Fathallah, N.B. Amor, P.~Leray, An optimized quantum circuit representation
  of bayesian networks, in \emph{European Conference on Symbolic and
  Quantitative Approaches with Uncertainty} (Springer, 2023), pp.
  160--171

\bibitem{RishOthers-411}
I.~Rish, Others, An empirical study of the naive bayes classifier, in
  \emph{IJCAI 2001 workshop on empirical methods in artificial intelligence},
  2001, vol.~3, pp. 41--46

\bibitem{Leung-429}
K.M. Leung, Others, Naive bayesian classifier, Polytechnic University
  Department of Computer Science/Finance and Risk Engineering, 2007, 123

\bibitem{Kononenko-450}
I.~Kononenko, Semi-naive bayesian classifier, in \emph{Machine
  Learning—EWSL-91: European Working Session on Learning} (Springer, Porto,
  Portugal, 1991), pp. 206--219

\bibitem{Zhou-423}
Z.H. Zhou, \emph{Machine learning} (Springer Nature, 2021)

\bibitem{Mindquantum-418}
M.~Developer.
\newblock Mindquantum, version 0.6.02021.
\newblock \urlprefix\url{https://gitee.com/mindspore/mindquantum}

\bibitem{LecunBottou-417}
Y.~LeCun, L.~Bottou, Y.~Bengio, P.~Haffner, Gradient-based learning applied to
  document recognition, Proc. IEEE, 1998, 86(11), 2278

\bibitem{XiaoRasul-416}
H.~Xiao, K.~Rasul, R.~Vollgraf, Fashion-mnist: a novel image dataset for
  benchmarking machine learning algorithms, arXiv preprint arXiv:1708.07747,
  2017

\bibitem{Offmann-RgensenPisier-451}
J.O.R. Offmann-J O~Rgensen, G.~Pisier, The law of large numbers and the central
  limit theorem in banach spaces, Ann. Probab., 1976, pp. 587--599

\bibitem{Camerini-452}
P.M. Camerini, The min-max spanning tree problem and some extensions, Inf.
  Process. Lett., 1978, 7(1), 10

\bibitem{NielsenChuang-404}
M.A. Nielsen, I.L. Chuang, \emph{Quantum Computation and Quantum Information}
  (Cambridge University Press, Cambridge, UK, 2000)

\bibitem{YuWang-424}
D.~Yu, H.~Wang, P.~Chen, Z.~Wei, Mixed pooling for convolutional neural
  networks, in \emph{Rough Sets and Knowledge Technology: 9th International
  Conference, RSKT 2014}, pp. 364--375

\bibitem{Dutilleul-453}
P.~Dutilleul, The mle algorithm for the matrix normal distribution, J Stat.
  Comput. Simul., 1999, 64(2), 105

\bibitem{BoullE-425}
M.~Boull~E, Modl: a bayes optimal discretization method for continuous
  attributes, Mach. Learn., 2006, 65, 131

\bibitem{Liu-408}
Y.~Liu, Z.~Yangming, A strategy on selecting performance metrics for classifier
  evaluation, Int. J. Mob. Comput., 2014, 6(4), 20

\bibitem{Caelen-407}
O.~Caelen, A bayesian interpretation of the confusion matrix, Ann. Math. Artif.
  Intell., 2017, 81(3--4), 429

\bibitem{GaussianNB}
Gaussian naive bayes.
\newblock
  \urlprefix\url{https://scikit-learn.org/stable/modules/generated/sklearn.naive_bayes.GaussianNB.html}

\end{thebibliography}
\end{document}